\documentclass[aps,prl,showpacs,amsmath,nofootinbib
,superscriptaddress,twocolumn,preprintnumbers]{revtex4}

\usepackage{natbib}
\usepackage{amsmath}
\usepackage{amssymb}
\usepackage{graphicx}
\usepackage{color}

\begin{document}

\setlength{\unitlength}{1mm}

\title{Scale Invariant Gravitation and Unambiguous Interpretation of Physical Theories}

\author{Meir Shimon}

\affiliation{School of Physics and Astronomy, 
Tel Aviv University, Tel Aviv
69978, Israel}
\email{meirs@wise.tau.ac.il}

\begin{abstract}
Our conventional system of physical 
units is based on local or microscopic {\it dimensional} 
quantities which are {\it defined}, for convenience or 
otherwise aesthetic reasons, to be spacetime-independent. 
A more general choice of units may entail variation 
of fundamental physical quantities (`constants') 
in spacetime. Whereas the theory of gravitation - 
indeed any theory of fundamental interactions 
minimally coupled to gravitation - is invariant to 
general coordinate transformations, it generally 
does not satisfy conformal symmetry, i.e. it is not 
invariant to local changes of the unit of length. 
Consequently, the {\it dimensionless} action associated 
with the Einstein-Hilbert action ($S_{EH}$) of 
gravitation, $\phi_{EH}=S_{EH}/\hbar$, is not invariant 
to local changes of the length unit; clearly an 
unsatisfactory feature for a dimensionless quantity.
Here we amend the phase by adding extra terms that 
account for spacetime variation of the physical 
`constants' in arbitrary unit systems. In such a unit 
system, all dimensional quantities are implicitly
spacetime-dependent; this is achieved by a conformal 
transformation of the metric augmented by appropriate 
metric-dependent rescalings of the dimensional quantities.
The resulting modified dimensionless action is 
scale-invariant, i.e. independent 
of the unit system, as desired. The deep connection 
between gravitation, dimensionless physical quantities, 
and quantum mechanics, is elucidated and the implicit 
ambiguity in interpretations of dimensional quantities 
is underlined. The emerging interpretation of the 
gravitational interaction is rather surprising; 
indeed, it provides a prescription to the dynamics 
of spacetime for a given energy-momentum distribution, 
as in the conventional view, but in addition it 
determines how the fundamental {\it dimensional} 
`constants'  - such as particle masses, electric charges, 
quantum of angular momentum - are all determined 
{\it locally} by the {\it entire} energy-momentum 
distribution in the universe. In an arbitrary unit 
system one cannot distinguish the dynamics of spacetime 
from the `dynamics' of the standard `rulers'; only the 
dynamics of dimensionless quantities is amenable to 
unique, unambiguous interpretation.
\end{abstract}

\pacs{03.65.Ta, 98.80.-k}

\maketitle

{\bf Introduction}.-- 
Over three centuries after Newton and a century after 
Einstein's seminal contributions to our understanding of 
the gravitational force, it remains enigmatic. 
A century later, Einstein's vision of unifying 
the fundamental physical interactions is still hampered 
by the defiant theory of gravitation; marrying 
gravitation with the fundamental principles of quantum 
mechanics proved to be an insurmountable task. 

A desirable feature of any physical theory is its 
background-independence; the laws of physics should 
be independent of the coordinate reference frame. 
Alternatively, physical laws should be 
independent of the state of the observer, as 
was first illustrated with Einstein's general 
relativity (GR). 
While being invariant under general 
coordinate transformations GR is 
not scale-invariant [1, 2], implying that Einstein equations 
(i.e. the energy-momentum conservation laws) apply 
in their standard form only when the 
conventional standard rulers 
of distance, time, and mass, are fixed to constant values by 
observations of the local universe or atomic physics.

Perhaps because of how experimental 
physics evolved, and possibly also due to 
prejudice,
local physical units have been fixed to constant values. 
This sets our standard length, mass, and time 
system of units (SU). 
By convention, all like particles have the same mass, 
electric charge, 
etc., the defining properties 
of elementary particles. 
However, all these properties are dimensional and therefore depend 
on the SU. A constructive example is from 
the field of cosmology. 
By measuring the cosmological redshift of light 
emitted by distant astronomical objects cosmologists have 
concluded that the universe is expanding [3-7]. This conclusion 
is further supported by theories of structure formation in 
an expanding background space which successfully explain 
a wealth of observed phenomena associated with the large 
scale structure [4-7]. However, {\it observations 
really only inform us that the dimensionless ratio 
of local and cosmological scales is a decreasing function of time.} 
This standard conclusion 
relies on choosing {\it local} length scales 
as our rulers, i.e. assuming they are fixed.

An equally valid choice of the standard length ruler
would be the distance between two remote galaxies [8]. 
Then, the universe is static on cosmological scales 
and consistency with observations requires that, e.g., 
microscopic scales (or even galactic scales) must 
contract over time. This conclusion may sound 
counter-intuitive but our intuition derives from 
arbitrary conventions that 
are motivated by 
convenience or due to historical reasons; an 
observer inside the solar system cannot rule out the 
possibility that everything in the system, 
including the observer himself, local ruler, and the 
subject of measurement, are all 
contracting with respect to the fixed distance 
between the galaxies. None of these two different 
interpretations is unit-independent; only statements 
about dimensionless quantities are unambiguous [9,10] 
and statements about cosmological expansion or contracting galactic 
scales (or Compton wavelengths for that matter) 
concern dimensional 
quantities and are therefore equally ambiguous. 

Of the four known fundamental interactions the 
gravitational force stands out as intimately related to 
the geometry of spacetime itself. Therefore, it appears in 
any consistent formulation of the other three physical 
interactions, via the covariant volume element 
and covariant derivatives. 
This feature, as we show in this {\it letter}, 
may be viewed as not only a mathematical manifestation of the 
general coordinate covariance of the fundamental interactions. 
Rather, it may also be viewed as 
the manifestation of the spacetime dependence of 
dimensional physical {\it quantities}, 
e.g. the speed of light $c$, the Planck constant $\hbar$, 
Newton's gravitational constant 
$G$, the electron charge $e$, and even particle masses, 
in an unconventional SU. 
This entirely different picture results in identical predictions 
to those of the conventional view; In a general 
SU, we argue, gravitation comprehensively 
captures both the metric dynamics and spacetime variation of 
length ruler.
From this perspective the dynamics of spacetime 
might be viewed (at least partially) as 
an emergent phenomenon associated with our prejudice 
that the fundamental physical quantities 
are indeed constant.
This wider symmetry than merely general 
coordinate invariance, is broken by adopting a fixed SU, 
the conventional system of units (CSU), in which the physical 
dimensional `constants' are truly constants.
The question then arises -- `constant' with respect to what ? 

Underlying our construction is quantum phase 
invariance, which is discussed in the next section.
A description of the physical picture in different 
metric frames is given in section 3. 
A few important ramifications are highlighted in 
section 4 and a brief summary is given in section 5. 

{\bf Invariance of Quantum Mechanical Transition Amplitudes}.--
The principles of quantum mechanics seem to 
fundamentally underly nearly every physical theory we know of. 
It is therefore desirable to incorporate them in our construction. 
We require the dimensionless action (in units of 
its quantum) to be invariant to SU changes, 
i.e. $\phi=S/\hbar$ is invariant. In arbitrary SU that allows 
the spacetime variation of $\hbar$ the phase 
reads $\phi=\int dS/\hbar$. The transition amplitude 
from a quantum state $\psi_{1}$ to $\psi_{2}$ 
is $\Gamma_{1\rightarrow 2}=\sum_{hist}\exp(i\int dS/\hbar)$ 
where the sum runs over all possible histories (trajectories) 
subject to fixed initial and final states, $\psi_{1}$ and 
$\psi_{2}$, respectively [11,12]. These trajectories can be 
defined either 
in real space [for a quatum point particle (PP)] or in configuration 
space (in case of quantum fields). In a manifestly 
generally covariant form the action reads 
$S\equiv\int\mathcal{L}\sqrt{-g}d^{4}x$ with $\mathcal{L}$ and 
$\sqrt{-g}d^{4}x$ the Lagrangian density (LD) and the covariant 
4-dimensional volume element, respectively, and $g$ is 
the determinant of the spacetime metric $g_{\mu\nu}$. 
The overall phase is unobservable but the 
relative phases associated with different trajectories 
are observable in interference experiments and, consequently, 
in the probabilities for quantum transitions. 
Two theories are considered 
equivalent if they result in identical probabilities 
for all possible quantum transitions/processes.
Two theories with equivalent phases $\phi=\int dS/\hbar$ 
will have equivalent transition probabilities, 
$P_{1\rightarrow 2}\propto |\Gamma_{1\rightarrow 2}|^{2}$.
Since the underlying physical laws are quantum 
mechanical and classical mechanics is at best an excellent 
approximation, any physical process/observable 
is governed by this transition amplitude.
Below, we construct a continuum of different US 
that result in exactly the same phase; while not 
unexpected, this highlights the relation between quantum 
mechanics and the irrelevance of spacetime variation 
of dimensional quantities.

{\bf Constructing the Correspondence Between Systems of 
Units}.--
In the following, the notion of transformation between 
SU refers to conformal metric rescaling 
augmented by rescaling of dimensional quantities.
The dimensionless action for a PP of mass $m$ in the CSU is
\begin{eqnarray}
\phi=\int dS/\hbar=\int (mc/\hbar)ds=
\int (mc/\hbar)\frac{ds}{d\lambda}d\lambda
\end{eqnarray}
where the infinitesimal interval is defined as
\begin{eqnarray}
ds^{2}=g_{\mu\nu}dx^{\mu}dx^{\nu}
\end{eqnarray}
and Greek indices run over all four spacetime dimensions, 
Einstein convention implied, and $\lambda$ is an 
affine parameter. 
If $mc/\hbar$ is spacetime-dependent the effective 
metric associated with the phase becomes 
$\tilde{g}_{\mu\nu}\propto (mc/\hbar)^{2}g_{\mu\nu}$.

It is useful to rewrite Eq.(1) in terms of the 
phase $\phi_{pp}=\int\mathcal{L}_{pp}\sqrt{-g}d^{4}x/(\hbar c)$
where $d^{4}x=c\cdot dtdxdydz$ in Cartesian coordinates, 
and the LD associated with a PP is
\begin{eqnarray}
\mathcal{L}_{pp}&=&\int mc
\sqrt{g_{\mu\nu}\dot{x}^{\mu}\dot{x}^{\nu}}
\frac{\delta^{4}[{\bf x}-{\bf x}(\lambda)]}{\sqrt{-g}}d\lambda
\end{eqnarray}
with ${\bf x}(\lambda)$ describing the particle trajectory 
and $\dot{x}^{\mu}\equiv\frac{dx^{\mu}}{d\lambda}$.

For a pure gravitational system the phase 
reads $\phi_{gr}=\phi_{pp}+\phi_{EH}$
where $\phi_{EH}=\int\mathcal{L}_{EH}\sqrt{-g}d^{4}x/(\hbar c)$ 
determines the metric as a function of the matter distribution, 
with the Einstein-Hilbert (EH) LD
\begin{eqnarray}
\mathcal{L}_{EH}=\frac{\hbar c}{l_{p}^{2}}(R-2\Lambda)+T,
\end{eqnarray}
where $l_{p}=\sqrt{G\hbar/c^{3}}$ is the Planck length, and $R$, $T$, 
and $\Lambda$, are the Ricci scalar, the 
trace of the energy momentum tensor $T_{\mu\nu}$, and the 
cosmological constant, respectively.

It is well known that the EH action, and the Einstein 
field equations in particular, are not scale-invariant, 
i.e. they transform nontrivially under the 
conformal transformation 
$g_{\mu\nu}\rightarrow\Omega^{2}g_{\mu\nu}$ where 
$\Omega({\bf x})$ is an arbitrary spacetime-dependent scalar 
function [1, 2]. In other words, the EH action 
is sensitive to local redefinition of the units of 
distance and time. This implies that  
$\phi_{EH}$ depends on the arbitrary SU although 
it is dimensionless. This must be an artifact 
of changing the unit of length in the EH action 
while holding the physical constants unchanged.
It is shown below that this units redefinition 
results in a scale-invariant $\phi_{EH}$ if the physical 
`constants' are allowed to vary in a certain 
manner.

For $\tilde{g}_{\mu\nu}=\Omega^{2}g_{\mu\nu}$ one has 
$\sqrt{\tilde{-g}}=\Omega^{4}\sqrt{-g}$ in four spacetime 
dimensions and the Ricci scalar transforms as 
$\tilde{R}=\Omega^{-2}(R-6\Box\Omega/\Omega)$, where 
$\Box f=\frac{1}{\sqrt{g}}\partial_{\mu}(\sqrt{g}
g^{\mu\nu}f_{,\nu})$ is the D'Alambertian of the 
function $f$ on a curved spacetime described by the 
metric $g_{\mu\nu}$ [13]. Let us assume that $\mathcal{L}$ 
is scale-invariant. This subclass of transformations is particularly 
interesting in the cosmological context [8]. 
In this case, $\tilde{T}=T$, i.e. 
the energy (E) per volume, must also be scale-invariant implying that 
$E\propto\Omega^{3}$. 
For the contribution of the matter sector to $\phi_{EH}$ 
to satisfy scale-invariance, $\sqrt{-g}d^{4}x$ must scale as $\hbar c$, 
and since $\sqrt{-g}d^{4}x$ has units $length^{4}$ each 
length dimension must scale as $\Omega$, i.e. 
$\tilde{l}_{p}=\Omega l_{p}$. Further 
requiring that time scales are scale-invariant, 
velocities must transform as lengths, and therefore 
$\tilde{c}=\Omega c$.
For $\phi_{EH}$ to be scale-invariant, $\tilde{\hbar}=\hbar\Omega^3$. 
Plugging this latter scaling in the transformation 
of $l_{p}$ results in the scaling of Newton's 
gravitational constant $\tilde{G}=G\Omega^{2}$. 
Requiring the Compton wavelength $\lambda _{c}=\hbar/(mc)$ 
scaling to match that of $l_{p}$ one obtains 
$\tilde{m}=m\Omega$. It is easy to verify 
that typical times for gravitational collapse, 
$t_{gr}=(G\rho/c^{2})^{-1/2}$, where $\rho$, the 
energy density, is scale-invariant. Finally, 
to guarantee the scale-invariance of `geometric contribution', 
i.e. Ricci scalar and cosmological constant, 
to $\phi_{EH}$ it is necessary to modify the EH LD to
\begin{eqnarray}
\frac{l_{p}^{2}}{\hbar c}\mathcal{L}_{EH}=
R-2\Lambda+6\Box f/f-12g^{\mu\nu}(\ln f)_{,\mu}(\ln f)_{,\nu}
\end{eqnarray}
where $f$ is any physical quantity (or combinations thereof) 
that scales as $\Omega$, e.g. $f\propto$ $c$, $l_{p}$, 
$\sqrt{e}$, $\sqrt{G}$, etc. In the CSU this reduces to 
$\frac{l_{p}^{2}}{\hbar c}\mathcal{L}_{EH}=R-2\Lambda$. 
The classical electron radius $r_{e}=e^{2}/(m_{e}c^{2})$ should 
scale as the Planck length, resulting 
in $\tilde{e}=\Omega^{2}e$, thereby illustrating that 
dimensionless quantities such as the fine structure constant, 
$\alpha=e^{2}/(\hbar c)$, are scale-invariant. 
Using conformal field theory 
terminology: angular momenta and energies, the coupling 
constants $G$ \& $e$, and distances and masses, are of 
conformal weights 3, 2, and 1, respectively. 
Dimensionless quantities have 0-weight.
Since $\alpha$ is scale-invariant the Rydberg `constant' 
$R_{\infty}=m_{e}c\alpha^{2}/(2h)\propto\Omega^{-1}$, 
scales as expected from a quantity with inverse 
length units.

Other scalings for $\mathcal{L}$ will result in 
different scalings of the physical `constants'. 
Assuming $\mathcal{L}\propto\Omega^{2\delta}$ and 
$c\propto\Omega^{\beta}$ where $\delta$ and $\beta$ 
are constants, and following similar arguments, 
the following scalings are obtained: 
$m\propto\Omega^{3+2\delta-2\beta}$, 
$\hbar\propto\Omega^{4+2\delta-\beta}$, 
$G\propto\Omega^{-2+4\beta-2\delta}$, and 
$e\propto\Omega^{2+\delta}$. Time scales as 
$\propto\Omega^{1-\beta}$. Since $\beta$ and $\delta$ 
are arbitrary, and furthermore $\Omega$ is an arbitrary 
function of spacetime, this represents a continuum of 
SU that leave the phase scale-invariant and consequently a 
continuum of possible interpretations of the 
same theory, if we insist on drawing conclusions 
from the dynamics of {\it dimensional} quantities. 
For concreteness we focus on the case $\delta=0$ 
and $\beta=1$ but the conclusions are general.

All this implies, that $\phi_{pp}$ is indeed scale-invariant. 
As a consequence, any observation involving the transition 
between quantum states in the presence of a pure gravitational 
field is scale-invariant, irrespective of the SU used in describing 
the process. 

Next, we incorporate the contribution from the 
electromagnetic (EM) interaction in the phase 
$\phi_{gr+EM}=\phi_{pp}+\phi_{EH}+\phi_{EM}+\phi_{EM.\ int.}$,
where the phases
\begin{eqnarray}
\phi_{EM}&=&\int F_{\mu\nu}F^{\mu\nu}
\sqrt{-g}d^{4}x/(\hbar c)\nonumber\\
\phi_{EM.\ int.}&=&\int\frac{e}{c\hbar} A_{\mu}dx^{\mu}
\end{eqnarray}
are associated with the free EM field 
and the interaction between the EM field and the EM 
current, respectively, and $F_{\mu\nu}$ is the EM 
field strength. From the fact 
that $A_{\mu}$ has units of $e x_{\mu}/r^{2}$ 
it can be readily seen that the interaction phase is, 
up to a dimensionless multiplicative constant factor, 
the fine structure constant, $\alpha$, and as 
illustrated above this quantity is scale-invariant. 
The combination $F_{\mu\nu}F^{\mu\nu}$ has units of 
$\mathcal{L}$ and is therefore scale-invariant, and since 
$\tilde{c}\tilde{\hbar}=\Omega^{4}c\hbar$, 
it immediately follows that the phase is indeed scale-invariant. 
Going beyond the PP approximation, one can repeat a 
similar procedure for the quantum electrodynamics 
(QED) phase, redefine the fields, masses, and charges 
and show that the phase is scale-invariant. 
This can be consistently done for the electroweak and 
strong interactions or any other 
contribution to the overall phase.

{\bf Ramifications}.--The idea of varying `constants' 
dates back at least to Dirac's proposal in 1937 that 
$G$ is a monotonically decreasing function of time 
and that in the remote past the gravitational and 
electric forces could have had a comparable strength 
[14]. Many other ideas along similar lines have been 
proposed over the years, e.g. Brans-Dicke scalar-tensor 
theory [15,16], varying electric charges [17], and 
varying speed of light [18-23], etc. These theories 
predict departures from standard physics -- departures 
which have been generally bounded by 
experiments/observations to marginal levels [24, 25].  
In contrast, the correspondence pointed out in this 
work between different SU interpretations, requires
all physical `constants' to vary simultaneously in 
a concerted manner that leaves 
dimensionless quantities invariant, 
the ratio between the strengths of the gravitational 
and EM fields in particular.

A seemingly odd feature of all SU other than CSU 
is the apparent non-conservation of, e.g., energy, 
linear momentum, angular momentum, and the electric 
charge. While general coordinate covariance is 
indeed a manifestation of energy momentum conservation, 
and the $U(1)$ gauge invariance of EM reflects electric 
charge conservation, they are obtained in CSU and by no 
means do they apply in a general SU.
In SU that allow the variation of fundamental 
`constants', Noether's theorem 
should be generalized and the conserved currents are 
associated with symmetries of the 
phase, i.e. the dimensionless action $\phi=\int dS/\hbar$, 
not the action $S$.

The ramifications that the SU correspondence described 
here might have on the interpretation of cosmological 
observations are studied elsewhere [8].
It is shown that the expanding universe interpretation with 
fixed physical constants can be consistently replaced 
with a static spacetime accommodating implicit 
time-dependence of the physical `constants', e.g. 
$c\propto a^{-1}$, $e$ and $G$ are $\propto a^{-2}$, 
$\hbar\propto a^{-3}$, and masses scale as 
$\propto a^{-1}$, where $a$ is the standard scale 
factor of the Friedmann-Lemaitre-Robertson-Walker 
(FLRW) spacetime [4-7], all this without altering the 
{\it predictions} of the conventional, dynamic spacetime, 
picture. 
The Einstein field equations serve to determine the 
`scale factor' $a$ as a function of time given the 
energy-momentum tensor. 
According to this viewpoint cosmological inflation 
[26, 27], as well as the {\it inferred} recent 
exponential expansion [28, 29], might be attributed 
to exponentially decreasing speed of light in a static 
spacetime [8]. Similarly, the {\it observed} 
cosmological redshift might be explained by the decreasing 
speed of light whereas the photon wavelength is constant. 
The relative strength of
the gravitational and EM, as well as 
the nuclear interactions, are fixed and therefore 
the physics of the early universe is unaltered. 

{\bf Summary}.--The physical laws are beautifully 
summarized in a set of four types of fundamental 
interactions that represent the very high symmetry 
of the universe. These symmetries are 
associated with a few fundamental conservation laws
which can be viewed as manifestations of the high 
symmetry of our universe, but it 
should be noted that these symmetries also reflect the 
CSU symmetry where dimensional quantities associated 
with atomic physics, e.g. particles masses, $\hbar$, 
and $e$ are fixed constants.
For example, the mass of 
the electron, $\hbar$, and $c$, 
can jointly define the system of units.
These standard rulers can well vary in 
spacetime and we have no direct observational evidence 
that this is not the case, nor can we have one, 
even in principle. It is actually 
meaningless to consider their variation since 
they are dimensional.
This may indeed be the case with {\it no observational 
imprint}, i.e. if gravitation is redundant with such 
local redefinition of the fundamental standard rulers 
then, the {\it dimensionless} phase, and more generally, 
quantum probabilities, cannot inform us about any 
such implicit spacetime variation of dimensional 
quantities.

As argued here, a range of physical phenomena, e.g., 
the interpretation of the expanding universe, distance 
and time measurements, etc., as well as
the `constancy' of fundamental physical quantities, such as 
$c$, $\hbar$, $e$, $G$ and fundamental particle masses, 
are all SU dependent. 
The idea that the observable physical world might have 
infinitely many viable `different' interpretations
may admittedly sound disturbing but, as explained, 
meaningful unambiguous interpretations apply only 
to dimensionless quantities, such as $\alpha$, 
redshifts, entropy, or probabilities 
for transitions between quantum states. 
This is analogous to the well-known 
fact that the metric associated with a given spacetime may 
or may not be singular depending on the 
coordinate system used. 
However, the real nature of spacetime is described 
by scalar quantities -- these are 
independent of 
the coordinate system. Similarly, different SU may 
result in different interpretations of the same 
physical reality unless dimensionless quantities 
are used. By generalizing the EH phase in 
Eq.(5) to arbitrary SU it is guaranteed that 
our description of physical reality is independent 
of arbitrary SU choices. The gravitational phase 
is then said to be scale-invariant. Conventionally, 
variation of this phase will result in the Einstein 
equations; given the energy-momentum tensor the 
metric is obtained. With Eq.(5), this can be done 
in any SU and the solution of the Einstein equations 
in an arbitrary system reflects not only the symmetries 
of the energy-momentum distribution and the coordinate 
system used, but also the `dynamics' of the dimensional 
physical constants; distinguishing the metric dynamics 
from the spacetime evolution of the rulers is generally 
impossible.

\end{document}